# Adversarial Machine Learning-Enabled Anonymization of OpenWiFi Data

Samhita Kuili [1], Kareem Dabbour [1], Irtiza Hasan [1], Andrea Herscovich [1],
Burak Kantarci [1], Marcel Chenier [2], and Melike Erol-Kantarci [1]

[1]School of Electrical Engineering and Computer Science, University of Ottawa, Ottawa, Canada
[2]NetExperience Inc., Ottawa, Canada
Emails: {skuil016, kdabb095, ihasa074, ahers063, burak.kantarci, melike.erolkantarci}@uottawa.ca
marcel@netexperience.com

*Abstract*—Data privacy and protection through anonymization is a critical issue for network operators or data owners before it is forwarded for other possible use of data. With the adoption of Artificial Intelligence (AI), data anonymization augments the likelihood of covering up necessary sensitive information; preventing data leakage and information loss. OpenWiFi networks are vulnerable to any adversary who is trying to gain access or knowledge on traffic regardless of the knowledge possessed by data owners. The odds for discovery of actual traffic information is addressed by applied conditional tabular generative adversarial network (CTGAN). CTGAN yields synthetic data; which disguises as actual data but fostering hidden acute information of actual data. In this paper, the similarity assessment of synthetic with actual data is showcased in terms of clustering algorithms followed by a comparison of performance for unsupervised cluster validation metrics. A well-known algorithm, K-means outperforms other algorithms in terms of similarity assessment of synthetic data over real data while achieving nearest scores 0.634, 23714.57, and 0.598 as Silhouette, Calinski and Harabasz and Davies Bouldin metric respectively. On exploiting a comparative analysis in validation scores among several algorithms, K-means forms the epitome of unsupervised clustering algorithms ensuring explicit usage of synthetic data at the same time a replacement for real data. Hence, the experimental results aim to show the viability of using CTGAN-generated synthetic data in lieu of publishing anonymized data to be utilized in various applications.

## I. INTRODUCTION

WiFi-enabled services have been booming with the advent of widespread adoption of wirelessly connected devices. Simultaneously, a potential vulnerability persists in ensuring the privacy of usage information. Previous studies employed data privacy techniques such as k-anonymity [1], l-diversity [2], t-closeness [3] and differential privacy [4]. However, those techniques can be complemented by considering data correlation, which possesses utmost significance while exploiting big data [5]. The heterogeneous characteristics underlying in OpenWiFi data possess identical traits of Big Data. Hence, adoption of novel machine learning techniques can eventually enable data privacy through anonymization.

Data anonymization is an essential task prior to making the data public for numerous domain applications. Apart from assuring privacy, it is also challenging to develop anonymized traces, which will resemble the original data. Therefore, retaining the resemblance of original in anonymized data by incorporating abundant noise alters the variance among features of data. This in turn raises issue on usability of data; which is where a trade-off exists while ensuring both data privacy and utility simultaneously on published data [6]. Previously, anonymization has been carried out mostly on healthcare-related domain pertaining to privacy for medical records of patients [7]. To the best of our knowledge, this paper paves the way for anonymization of WiFi usage streams for the first time.

A collaboration of unsupervised and generative adversarial network (GAN) is considered a holistic approach to initiate privacy preservation. Several clustering algorithms, namely K-means, Density Based Spatial Clustering of Application with Noise (DBSCAN) [8], Gaussian Hidden Markov Model [9] and Agglomerative are employed to analyze the actual pattern distribution of the usage traffic. While examining real-time data, there are no prior information on labels corresponding to each stream of data per timestamps. Therefore, employing clustering techniques provides useful insights to comprehend the behavioural patterns among samples in a multivariate data set. On the other hand, adopting a deep neural generative model GAN [10], [11] harbors generation of synthetic samples.

Commonly GAN is quite well-known in computer vision [12], [13]. There are other robust architectures for GANs which can yield remarkable performance on tabular data sets or non-image data sets. In light of these, this work aims to generate synthetic samples by leveraging a conditional tabular GAN (CTGAN) [14]; which is a modified version of architecture to traditional tabular GAN (TGAN) [15]. Hence, leveraging both clustering algorithms and computational power of CTGAN, we narrow down our work to anonymization by testing and validating in terms of quality of clusters. Therefore with support of cluster metrics it becomes easier to validate the performance of CTGAN; which aids us in assessing the quality of generated synthetic samples. At the same time, it is crucial to test the amount of distortion being carried out in original samples. Simultaneously, we want to ensure that the synthetic samples resembles statistical properties of the original data. Fig. 1 showcases topology for demonstration of data anonymization.

Key contributions of this paper are listed below with the ultimate goal to successfully build data anonymization while maintaining data utility and data privacy:

- The initiation of anonymization is being undergone by leveraging heuristic clustering algorithms: K-means, DB-SCAN, Gaussian Hidden Markov Model (GHMM) and Agglomerative on a normalized data to showcase the potential cluster labels from the standpoint of distance, density and probability function.
- Leveraging unsupervised algorithms to present the formation of cluster labels and considering clustering membership information as a discrete variable and one of the pivotal parameters to train CTGAN.
- In addition, production of synthetic samples by training CTGAN on the real normalized samples and adopting effective measures to validate similarity performance of synthetic over real samples in terms of cluster metrics, for instance, Silhouette, Calinski and Harabasz (CH) and Davies-Bouldin (DB) scores.

The aforementioned state-of-the-art aligns to aim for high-quality data anonymization technique while advocating for the reliability and trustworthiness of using synthetic data from the standpoint of best-unsupervised algorithm and a number of cluster validation scores regardless of the distorted distribution of data. Over the course of this study, meticulous background information is provided in Section II highlighting the importance of data privacy algorithms and details of use-cases for generative adversarial networks (GANs) in terms of the generation of synthetic data. Moreover, Section III discusses the coherency of the data anonymization undertaken gradually on a multivariate time series OpenWiFi data. Furthermore, an intensive evaluation showcases the significant performance comparison of adopted unsupervised clustering algorithms with regard to several validation metrics in Section IV, evaluating the quality of anonymized data in comparison to that of the real data. Finally, Section V provides the concluding remarks summarizing the research undergone in this study.

## II. RELATED WORK

Data anonymization is defined as the strategy of encrypting sensitive information specifically personally identifiable data in a dataset. In addition, the goal is to assure privacy protection of any information belonging to an attribute in the data by preventing information leakage when being exploited by others performing numerous use-cases. Among various methods of anonymization, for instance, generalization, masking, suppression, perturbation, and usage of synthetic data, the latter is given the major importance by advocating a machine learning model such as a Conditional Tabular Generative Adversarial Network (CTGAN). Consequently, prominent divergence infused on the real data reveals statistical alteration of the data while ensuring commonalities between synthetic and real data.

### A. Data Privacy Algorithms

Previously data privacy preserving methods such as $k$-anonymity, $l$-diversity, $t$-closeness and $\delta$-disclosure are leveraged to initiate privacy by compromising certain amendments to the attributes present in data set. Before exploring the process of those methods, it is crucial to understand following key terminologies identifiers, quasi-identifiers (QIDs) and sensitive attributes [16]. Generally, key identifiers signify those attributes with a unique code or number; which are lacking by default in our data. However, quasi-identifier signifies attributes with discrete value denoting multiple parameters; such as MAC address, location IDs, equipment IDs, and timestamps in our work. Lastly, sensitive attributes are those consisting of other attributes which can neither be termed as identifiers or quasi-identifiers (QIDs). Sweeney [17] proposed an algorithm k-anonymity by elaborating on data; associated with person specific information to ensure privacy of identity and simultaneously considering the issue of re-identification attack into account. In fact, this algorithm works suitably well for selective attributes. Therefore, there is always a way for launching an adversarial attack using other attributes which are not being anonymized using k-anonymity. k-anonymity performs poorly in term of guaranteeing privacy due to presence of unequivocal homogeneity and background knowledge attacks. This is exactly when l-diversity plays a humongous role in ensuring privacy. On the contrary, the adversary seems to develop a thorough knowledge of background distribution of data even after relying on l-diversity algorithm. In order to avoid this consequence, Li et al. [2] introduced a novel algorithm t-closeness, which aligns the distribution of sensitive attributes with other attributes and therefore diminishing plausible background knowledge. Differential privacy is another well-known technique which shows improvement while incorporating privacy measures related to health care domains. Jain et al. [4] discusses on the amount of distortion incorporated by database into the data; which is determined in maintaining privacy and consequently can be found useful by data analysts.

### B. Conditional Tabular Generative Adversarial Network

The above mentioned techniques still lack non trivial knowledge to convince data privacy for big data. As discussed earlier, OpenWiFi follows 5'Vs of big data and continual deployment of diverse data streams in high volume, therefore, those techniques are not sufficient to ensure data anonymization. Generative adversarial networks [11] are receiving notability lately for creation of fake synthetic samples and are employed for anonymization application by researchers. The architecture of this generative model comprises of two components: generator and discriminator. Those components can be designed by adopting different types of neural networks based on the applicability. Due to the complex architecture of GANs, classifying a sample into fake or original simply eases the entire process. This is the adversarial minmax game where the terminology of GANs turns up. The idea behind minmax game demonstrates that generator will minimize its performance of generating new synthetic training samples, on the contrary, discriminator will aim for maximizing its performance in classifying the generated and original samples into accurate scalar label.

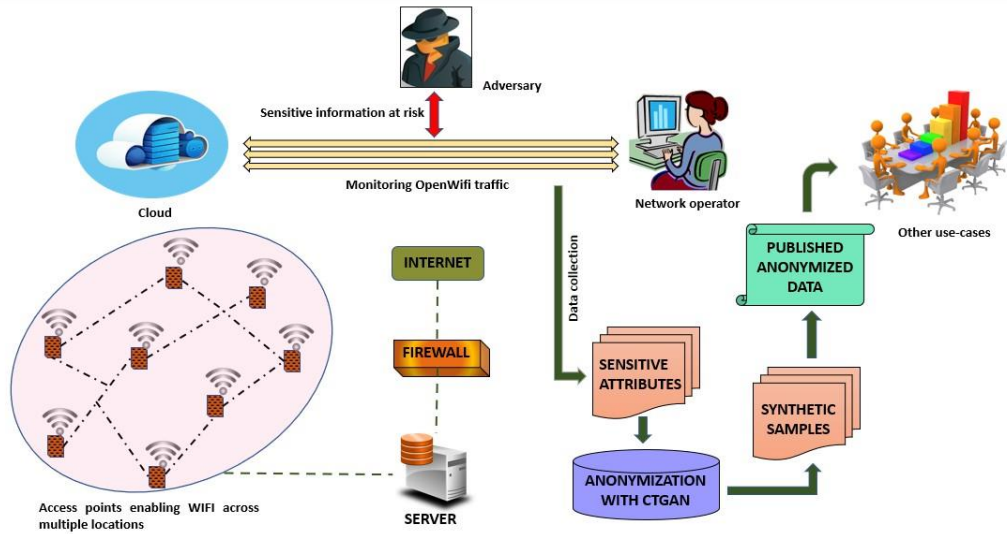

Fig. 1: Anonymization of OpenWiFi network traffic data.

$$\min_G \max_D V(D, G) = \mathbb{E}_{x \sim p_{data}(x)}[log(D(x))] + \mathbb{E}_{z \sim p_z(z)}[log(1 - D(G(z)))] \quad (1)$$

$p_z(z)$ is input noise distribution variable representing generator's distribution $p_g$ over data samples $x$. A data mapping $G(z;\vartheta_g)$ denotes differential function $G$ with parameters $\vartheta_g$. Similarly, there is another differential function $D$ with $\vartheta_d$ as parameters, which represents a mapping space for discriminator $D(x;\vartheta_d)$. In addition, $D(x)$ signifies likelihood of $x$ coming from real samples and not from distribution $p_g$. While training GANs, the motive is to train discriminator in assigning accurate labels for samples coming from each distribution mapping space. At the same time, train the generator to minimize the variable $log(1-D(G(z)))$. More clarification about the variables referred to (1) can be understood in [11].

There are other variant of generative models which are exploited for use case of generation of synthetic samples. Park et al. [16] introduce the adoption of tableGAN; which includes the architecture of deep convolutional generative adversarial networks (DCGAN) [12]. The main focus is being shown on generating relational synthetic tables based on existence of data attributes associated with convolutional neural networks [16].

Another interesting study undergone by Hajihassani et al. [18] which highlights their major contribution on anonymizing time series sensor data obtained from IoT devices with the help of Variational Autoencoder (VAE). Additionally, their work shares knowledge on understanding the anonymization into two aspects: algorithmic and systemic solutions. Considering algorithmic solutions involving machine learning, anonymization is being previously explored on the factors of security and protection with the assistance of deep neural networks and PrivacyNet [19]. Further insights on systemic solutions can be referred on [18] since those are not related with this work. As Xu and Veeramachaneni [15] propose the novel architecture of conditional tabular GAN emphasizing on the marginal analysis of data with an adoption of recurrent neural network, their work also shares other ways to obtaining synthetic data.

Adoption of other GANs, for instance, RGAN and RCGAN [20] are popular for generation of time series data. McCoy in [21] share the importance of CTGAN while training recommender systems which are useful to preserve user's privacy. Furthermore, this work also focuses on the reduction of actual raw training samples and in fact, expect an augmentation of actual samples in disguise of synthetic samples. Therefore, reliance on large retrieval of raw data is significantly reduced while still ensuring user privacy.

### III. METHODOLOGY

Prior to the experiment on anonymization, processing the raw data forms primary step of pipeline. This dataset comprises of 20,000 service records collected from a small network operated over 4 weeks. Moreover, this network is composed of 2 access points averaging 2 connected clients a day. Furthermore, these access points are up and running efficiently, with clients in ON/OFF mode. By setting the desired duration and timestamps, 20K records are collected via REST APIs Cloud Service Portal for OpenWiFi until sufficient information on the active access points and clients connected to them is retrieved. In terms of processing and setting raw data ready for CTGAN, removal of any specific quasi-identifiers; including but not limited to any equipment IDs, location IDs, or customer IDs are implemented. In addition, scaling raw data samples address concerns with mode-specific normalization of CTGAN. Thus, modelling a CTGAN builds on existing numerical or sensitive attributes that contain at most 15,000 traffic records forming a multivariate time series data set.

Moreover, collected data does not contain any prior information on labels. Therefore, adoption of unsupervised learning blends perfectly to comprehend underlying behavioural distribution of sensitive attributes. Those distributions are analyzed over clustering techniques such as K-means, Density Based Spatial Clustering of Application with Noise (DBSCAN) [22], Gaussian Hidden Markov Model (GHMM) [9]; establishing three different outlooks with respect to distance, density and probability factors. Moreover, an aggregation of clusters is visualized by selecting Agglomerative clustering algorithm. Quality of cluster validation for each unsupervised algorithm is assessed via Silhouette, CH and DB scores. At each stage of execution, potential clusters are demonstrated via Principal Component Analysis (PCA)-based dimensional reduction for more coherent understanding of original observable samples as well as synthetic samples.

Our processed multivariate data set consists of 15 feature variables with (11,900+) instances. While selection of optimal number of clusters for k-means clustering is challenging, setting a range from 2 to 10 is chosen. A trial and error execution is implemented to determine a best optimal number of cluster out of the arbitrary selection the aforementioned range. To obtain the optimal value k, a cluster validation metric Silhouette score is determined for each value k. Consequently, the value k for k-means turns out to be 2 with the highest Silhouette score ensuring grouping of observations based on distance heuristic with the focus on statistical similarity in data distribution of a multivariate dataset. Furthermore, the Silhouette score is an unsupervised clustering metric which quantifies the quality of the clustering of samples into the chosen optimal number of clusters i.e. k. The higher the Silhouette score (with a maximum score of 1) indicates a higher quality of grouping within each cluster and sufficient separability between them.. On the other hand, estimating optimal parametric values for DBSCAN (eps and minPts) undergoes an exhaustive search approach highlighting highest Silhouette score [22] The parameters *eps* and *minPts* are data dependant where *eps* stands for the radius to be followed with neighboring points from a particular data point in a cluster. In addition, *minPts* signifies the minimum neighboring data points required to be at a distance *eps* forming a dense distribution of data in a cluster. DBSCAN further identifies some samples as outliers which is bound to occur on high dimensional heterogeneous real data.

Therefore, minimum neighbouring points (minPts) 5 and a distance threshold epsilon (eps) 0.038 are selected. The number of states obtained for GHMM is domain-dependent. Kullback Leibler (KL) divergence has been used to obtain empirical probability distribution functions on HMM models with states 2, 3, and 4 on 1000 real RSSI WiFi signal samples [23]. Since there is a small difference between the KL divergence when using 2, 3, or 4 hidden states [24], a GHMM with 3 states is used to model this time series data. Additionally, in order to obtain a more cohesive view of the data, using agglomerative clustering with Ward linkage is also carried out in this work. Generally, agglomerative clustering follows an iterative hierarchical process starting from a single data point as a cluster and eventually merges into another cluster by identifying the closest pair of clusters.

## IV. EXPERIMENTAL RESULTS

Tuning of parameters to achieve the highest Silhouette score enables analysis of the points exhibiting similar statistical property grouped into clusters. Initially, the original behaviour of real data set right after preprocessing is learnt through clustering techniques as demonstrated in Fig. 2 (a-d); representing K-means, DBSCAN, GHMM and agglomerative algorithms respectively. On examining the Silhouette scores obtained for each technique, it is observed K-means acquires the highest Silhouette score with optimal cluster value 2 (k). Similarly, optimal cluster values for DBSCAN, GHMM and agglomerative obtains k = 3 respectively. At the same time, training a complex deep neural network architecture like CTGAN takes longer than expected while searching for best parameters suited for trained model. It is also challenging to validate the performance of synthetic samples generated by CTGAN. A well trained CTGAN is analyzed by understanding the performance loss convergence corresponding to two neural networks: generator and discriminator respectively as shown in Fig. 3.

Most common problems which thrive while training GANs are mode collapse and loss convergence. This incorporates adversarial game between two neural networks where either there is a decrement in performance of generator loss and simultaneously an increment on discriminator loss. In addition, initial performance on acquiring the cluster validation scores for real data is analyzed as in Table I. Furthermore, CTGAN is being trained to generate synthetic samples on being conditioned to discrete column of a data set; which contains cluster labels from each clustering algorithms at a time. The main motive is not limited to simply generate synthetic samples by exploiting generative model; but to check whether synthetic data can be relied upon as compared to real data. In order to assess the quality of synthetic data, clustering algorithms are employed which reflect the performance score for each synthetic and real data. As a result, more similarity or close the value for each metric, higher is the indication of replacing real with synthetic data. The initial real data (D) in Fig. 4 is set with target attribute; which consists of labels obtained from K-means.

Moreover, we retain dimensionality reduction of real data (D) through PCA obtaining low dimensional data (d); which also mitigates the problem collinearity among features. The core task for training generator of CTGAN requires conditional vectors. However, target attribute with labels 0 and 1 of K-means is considered as discrete variable while continuous variables as rest of the attributes or features existing on real data. Therefore, labels in discrete variable are utilized as conditional vectors to generate synthetic samples. A synthetic data ($D'$) is obtained after training CTGAN while tuning for convergence with specific hyper-parameters demonstrated in the form of a loss curve in Fig. 3 (a).

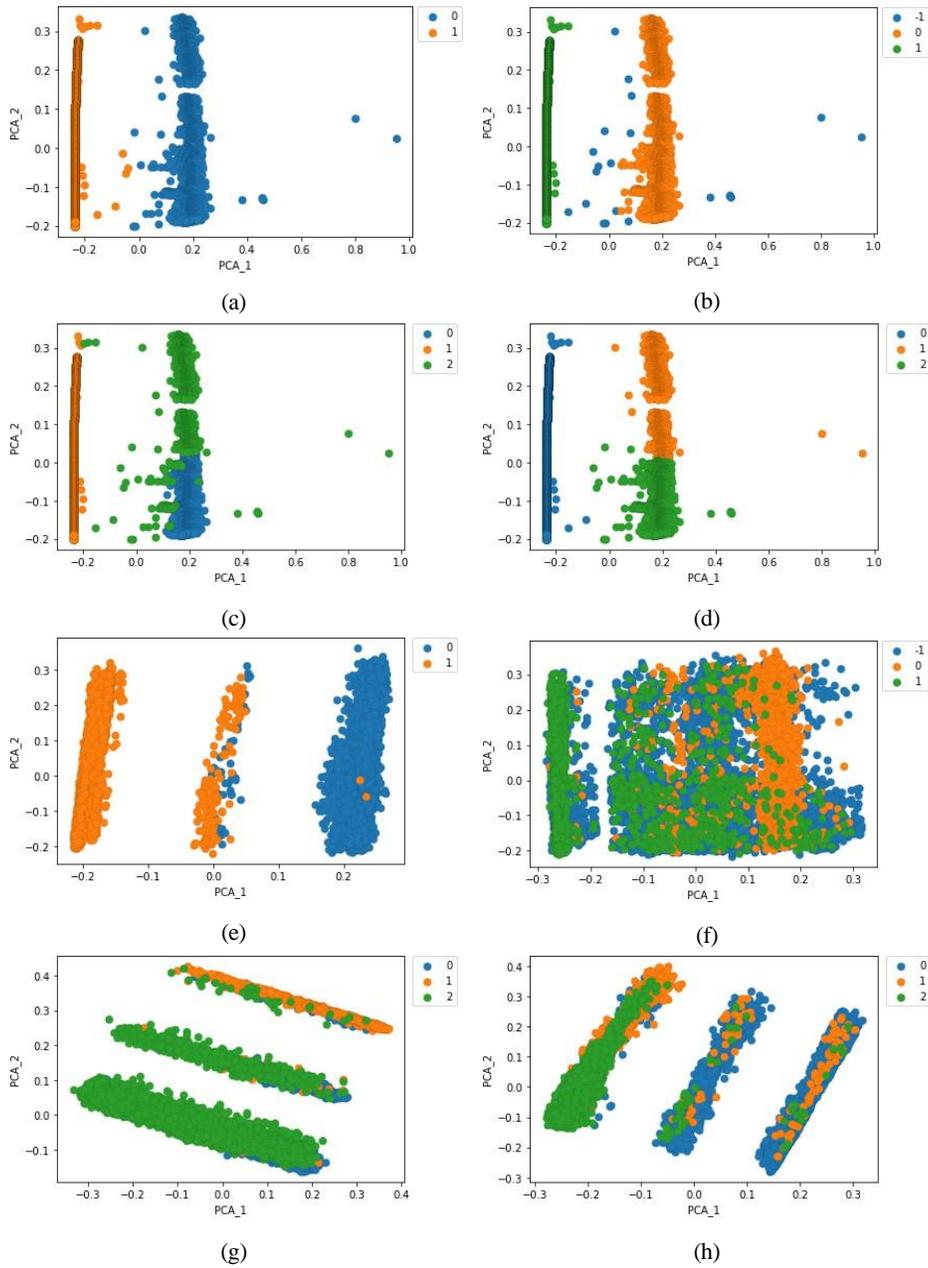

Fig. 2: Clustering visualization of real: (a-d) and synthetic data: (e-h).

This synthetic data ($D'$)) is further visualized with the help of PCA; forming another low dimensional data set ($d'$). Accordingly, evaluation of cluster metrics on ($d'$) are computed. This exact method is carried out followed by altering the target variable subsequently with labels of DBSCAN, GHMM and agglomerative clustering algorithms, and correspondingly scores are calculated. Consequently, loss curves demonstrated in Fig. 3 (b), (c) and (d) are also checked for convergence after setting the labels for DBSCAN, GHMM and agglomerative algorithms as conditional vectors for generator. The cluster validation scores computed on synthetic data can be understood from Table II.

While comparing the validation scores of K-means on both Table I and Table II, there is decent clarity which shows resemblance in scores for all three metrics. However, there is significant amount of deviation in scores for DBSCAN with respect to Silhouette, CH and DB. Similar to the Silhouette score, CH and DB scores aim to quantify quality of cluster formation. For instance, a higher CH score signifies better quality of clustering from the perspective of factors inclusive of within-cluster variance and between-cluster variance. Moreover, DB is quite the opposite of Silhouette and CH scores as it measures the average computation of inter-cluster distance and within-cluster dispersion of a cluster. More importantly, a lower DB score denotes greater cluster formation quality. While comparing scores for GHMM, CH score reflects high deviation as compared to Silhouette and DB. On the contrary, agglomerative achieves reasonable scores for all metrics of both real and synthetic data. In this study, we intend to look for similarity between the scores for both real and synthetic data. For instance, K-Means preserves the statistical properties in synthetic data even after incorporating noise or conditional vectors by CTGAN into real data.

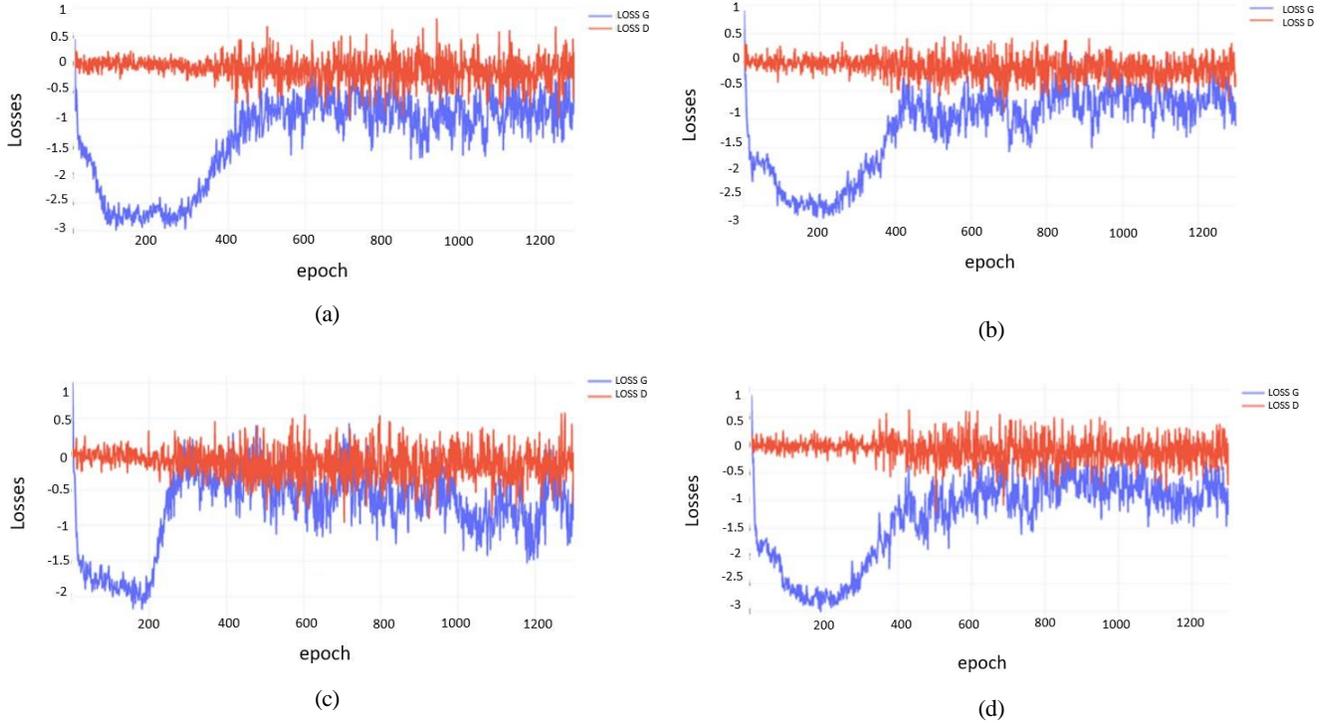

Fig. 3: Convergence of loss curves for CTGAN.

K-means is compatible while analyzing the performance of cluster validation scores from the perspective of data anonymization [25]. Additionally, a baseline algorithm K-Means on real data is selected to compare the cluster validation scores with each algorithm from synthetic data. It is observed that K-means exhibits resemblance in scores while comparing cluster validation metrics of real and synthetic data as shown in Table I and Table II, respectively. This coherently indicates that there is no significant statistical implication in the application context of data anonymization. However, DBSCAN, GHMM, and agglomerative algorithms result in the variation of both low Silhouette and Calinski Harabasz scores, and higher Davies Bouldin scores. In the context of application data anonymization and deviation in scores, it is understandable that DBSCAN is sensitive to outliers which results in performance downgrade under synthetic data when compared with the baseline algorithm. Similarly, the poor performance of GHMM and agglomerative algorithms highlights presence of noise in the synthetic data and involvement of model complexity with parameter initialization based on the data distribution. To comprehend deviation, we establish a visual analysis forming clusters of synthetic data for each algorithm shown in Fig. 2 (e-h); denoting K-means, DBSCAN, GHMM and agglomerative algorithms. On examine those figures, we encounter a skewed distribution of cluster labels as compared to Fig. 2 (a-d); which is rational to occur on a synthetic data. The reason of such distortion depends on the conditional vectors provided to train the generator ofs CTGAN. The resemblances in validation scores as seen on both real and synthetic data also provide assurance of preserving statistical properties similar to samples of real data; which is an essential factor as discussed in the earlier sections. Furthermore, this ensures that the synthetic data can be relied upon and effectively utilized for numerous applications.

TABLE I: Cluster validation scores of unsupervised algorithms on real data

| Unsupervised algorithms | Real data | | |
|---|---|---|---|
| | Silhouette | Calinski Harabasz | Davies Bouldin |
| K-means | 0.642 | 23342.92 | 0.599 |
| DBSCAN | 0.552 | 11659.34 | 5.559 |
| GHMM | 0.629 | 24788.87 | 0.529 |
| Agglomerative | 0.635 | 26109.21 | 0.501 |

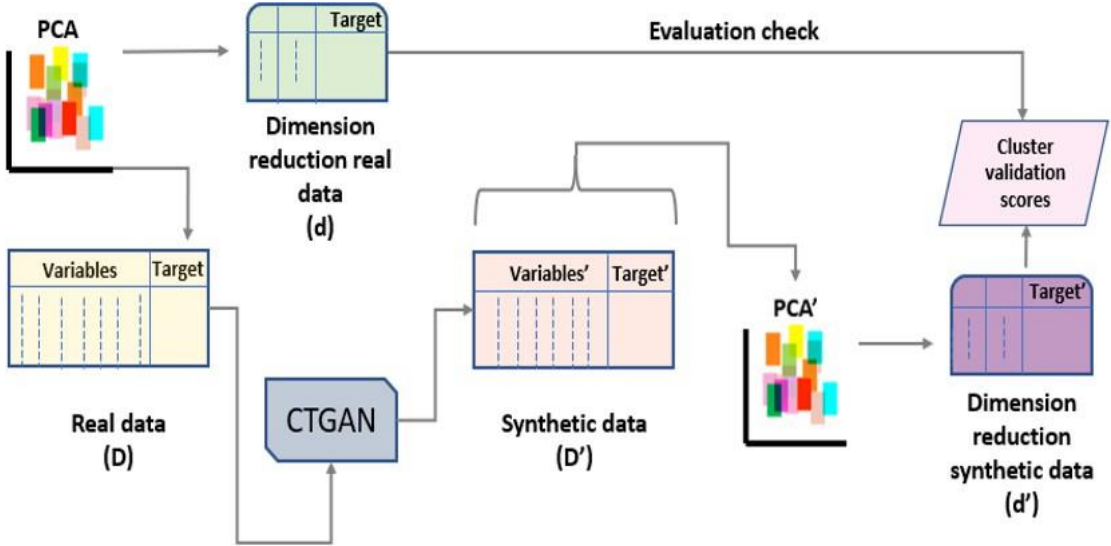

Fig. 4: Initiation of data anonymization leveraging clustering algorithms.

TABLE II: Cluster validation scores of unsupervised algorithms on synthetic data

| Unsupervised algorithms | Synthetic data | | |
| --- | --- | --- | --- |
| | Silhouette | Calinski Harabasz | Davies Bouldin |
| K-means | 0.634 | 23714.57 | 0.598 |
| DBSCAN | 0.171 | 3742.45 | 1.836 |
| GHMM | 0.409 | 8865.44 | 0.770 |
| Agglomerative | 0.507 | 12188.80 | 0.769 |

## V. Open Issues, Challenges, and Opportunities

The usage of multivariate time series OpenWiFi data where this study aims to adopt data anonymization and forecast the generation of the skewed distribution of synthetic samples already lack prior information of ground truth. Simultaneously, the analysis focuses on mimicking cluster labels / ground truth based on the data distribution present in real and synthetic data setting an example to comprehend the problem only from the standpoint of the statistical context. However, other ways to augment the trustworthiness of synthetic data include privacy risk assessment, evaluating downstream tasks, and incorporating the review by domain experts who are knowledgeable about real data and will provide valuable feedback on synthetic data.

Every new data anonymization technique needs to cope with several challenges. Re-identifying instances of anonymized data, and safeguarding data utility is just one to mention. The latter can be understood if there is a loss of pivotal information in the real data, which can pave the way for an effective decision capability while exploiting the anonymized data. In addition, the dynamicity of multivariate time series data needs to be taken into account since it consists of temporal dependency and may impact anonymized data. Moreover, the scalability of the proposed framework is also of paramount importance which needs to be addressed while dealing with large volume of data. However, consistent performance of data anonymization is expected while safeguarding the data utility regardless of exploiting big data. Furthermore, scalability may also be studied from the use of CTGAN, which expects a large volume of data to train itself, therefore would ensure better prediction ability by studying the distinctions between synthetic and real data.

The idea of adversarial machine learning-enabled anonymization imparts the efficacy of data sharing across different domains such as ethical use of data by adopting a measure of privacy protection of sensitive information in the data.

## VI. Conclusion

Data anonymization is a crucial task for operators to augment privacy of information; which furthermore meets required user demands. This work summarizes the generation of synthetic samples by deploying conditional tabular GAN to replace real data instances with synthetic samples. Incorporation of clustering mechanism highlighted heterogeneous distribution of OpenWiFi traffic on the grounds of underlying factors such as distance, density and probability of samples existing in both real and synthetic data. In addition, computation of cluster validation metrics by well-known Silhouette, Calinski and Harabasz and Davies-Bouldin scores have been undergone to comprehend reliability of synthetic data. A wise similarity comparison of those scores has been checked while assessing synthetic with real data. Moreover, two dimensional visualization at every stage of implementation enables us to understand the original and the skewed distribution of synthetic data. This work has particularly focused on resemblance in the unsupervised validation metrics, which invokes the preservation of statistically correlated properties among samples. By leveraging this knowledge, extension of this work includes incorporation of other machine learning and privacy-based metrics to improve data anonymization.

ACKNOWLEDGMENT

This work was supported in part by Mathematics of Information Technology and Complex Systems (MITACS) Accelerate Program under Project Number IT24702.